
\newskip\oneline \oneline=1em plus.3em minus.3em
\newskip\halfline \halfline=.5em plus .15em minus.15em
\newbox\sect
\newcount\eq
\newbox\lett

\def\tremat#1#2#3#4#5#6#7#8#9{\left(\matrix{#1&#2&#3\cr#4&#5&#6\cr
#7&#8&#9\cr}\right)}

\def\p32{{\tilde +}}
\def\m32{{\tilde -}}

\def\simlt{\mathrel{\lower2.5pt\vbox{\lineskip=0pt\baselineskip=0pt
           \hbox{$<$}\hbox{$\sim$}}}}
\def\simgt{\mathrel{\lower2.5pt\vbox{\lineskip=0pt\baselineskip=0pt
           \hbox{$>$}\hbox{$\sim$}}}}

\newdimen\short
\def\adv{\global\advance\eq by1}
\def\set#1#2{\setbox#1=\hbox{#2}}
\def\nextlet#1{\global\advance\eq by-1\setbox
                \lett=\hbox{\rlap#1\phantom{a}}}
\def\ov{\overline}
\newcount\eqncount
\eqncount=0
\def\equn{\global\advance\eqncount by1\eqno{(\the\eqncount)} }
\def\put#1{\global\edef#1{(\the\eqncount)}           }

\def\np{{\it Nucl. Phys.}}
\def\pl{{\it Phys. Lett.}}
\def\pr{{\it Phys. Rev.}}

\def\cPre{1}
\def\cSSb{2}
\def\cBac{3}
\def\cSS{4}
\def\cAnt{5}
\def\cAqm{6}
\def\cAB{7}
\def\cABQ{8}
\def\cx{9}
\def\cHet{10}
\def\cNar{11}
\def\cGin{12}
\def\cHig{13}
\def\corb{14}
\def\cWil{15}
\def\crank{16}
\def\cFI{17}
\def\cdtrm{18}

\magnification=1200
\hsize=6.0 truein
\vsize=8.5 truein
\baselineskip 14pt

\nopagenumbers

\rightline{hep-th/9509115}
\rightline{IC/95/306}
\rightline{September 1995}
\vskip 1.0truecm
\centerline {\bf PERTURBATIVE SUPERSYMMETRY BREAKING IN ORBIFOLDS}
\centerline{\bf WITH WILSON LINE BACKGROUNDS }
\vskip 1.0truecm
\centerline{{\bf Karim Benakli}
\footnote{$^*$}{\it e-mail: benakli@ictp.trieste.it}}
\vskip .5truecm
\centerline{\it International Centre of Theoretical Physics}
\centerline{\it  Trieste, Italy}

\vskip 2.5truecm
\centerline{\bf ABSTRACT}
\vskip .5truecm

A way to break supersymmetry in perturbative superstring
theory is the string version of the Scherk-Schwarz mechanism. There,
the fermions and bosons have mass splitting due to different
compactification boundary conditions. We consider the implementation
of this mechanism in abelian orbifold compactifications with Wilson
line backgrounds. For $Z_N$ and $Z_N\times Z_M$ orbifolds, we give the
possible $U(1)$ R-symmetries which determine the mass splitting, and
thus, the supersymmetry breaking at the perturbative level. The
phenomenlogical viability of this mechanism implies some dimension(s)
to be as large as the TeV scale. We explain how the lighter
Kaluza-Klein states associated with the extra-dimension(s) have
quantum numbers depending on the Wilson lines used.

\hfill\break

\vfill\eject

\footline={\hss\tenrm\folio\hss}\pageno=1

One of the standing questions of high energy physics is the presence
or the absence of higher (spontaneously broken) symmetries of
nature beyond the ones of the standard model. Among the
expected or wanted ones is spontaneously broken supersymmetry.
Supersymmetry appears quite naturally in superstring theory. However
understanding its breaking directly at the level of perturbative
four-dimensional superstring theory remains up to now an open problem.
Previous studies [\cPre,\cSSb], as well as a more recent
attempt [\cBac], have all lead to the prediction that at least one of
the internal dimensions needs to be large. More precisely, because of
the non-renormalization theorems, supersymmetry has to be broken at
tree level. Then the gravitino or gauginos get masses inversely
proportional to the size of some internal dimension.
As these masses are usually required to be of order of the
electroweak scale in order to protect the gauge breaking scales
hierarchy, it results that the size of some
internal dimension(s) is of the order of the TeV.

A possible way to build theories with
broken supersymmetry is the string version of the
Scherk-Schwarz mechanism [\cSS]. Such theories have become recently of
some phenomenological interest after it has been shown that they
could allow a weakly coupled string theory, at least at one-loop for a
class of models based on orbifold compactifications [\cAnt]. This
result also suggests that the manifestation of the large
extra-dimension(s) would be the existence of some new states with
regularly spaced masses and behaving as excitations of the MSSM
particles [\cAqm]. In the limit where some supersymmetry (thus
electroweak) breaking effects are neglected, the quantum
numbers and interactions of these states have been investigated in
[\cAB]. There, it has been pointed out that in a minimal
scenario, their only observable effects are through some
non-renormalizable effective operators. The latter have been computed
and limits on the size of new dimensions have been derived from actual
experimental data [\cAB]. The obtained bounds allow the hope of
experimental detection in the near future [\cABQ].

The Scherk-Schwarz mechanism of supersymmetry breaking has been up to
now analyzed only for string models with large gauge groups obtained in
the absence of any gauge background fields. We will extend this
analysis to the case where Wilson lines are present.
These are necessary to break the former large groups into
smaller ones like $SU(3)\times SU(2)\times U(1)$ at the string
level. We will first introduce the formalism which allows us to
derive the mass spectrum in presence of both Wilson lines  and
supersymmetry breaking. We will notice that the two effects can be
studied separately. One of our results will be to give the
possible charges that could be used in orbifold models to break
supersymmetry \`a la Scherk-Schwarz in the context of ${\bf Z}_N$ and
${\bf Z}_N \times {\bf Z}_M$ orbifolds. We will then turn
to the study of some important properties of the Kaluza-Klein
(KK) states in the presence of Wilson lines. Although, the spectrum in
absence of supersymmetry breaking is well known, the implications
for the case of a dimension of very large size haven't been studied.
One of our results will be to show that the KK excitations form
(spontaneously broken) $N=4$ or $N=2$ multiplets with quantum numbers
depending on the choice of the Wilson lines. In particular, if
these are associated to the large dimension(s), then new very light
states are present with quantum numbers which cannot be expected only
from the knowledge of the massless spectrum. But we will argue that
the rank of the KK symmetry group remains the same as the one of the
(massless) gauge group if the Wilson lines are all associated with the
other  dimensions of small size. Some attempts to study explicit
orbifold models with minimal content of KK states will be presented
elsewhere [\cx].

Let us first derive the spectrum of the KK excitations in presence
of both  Wilson lines and supersymmetry breaking. The
worldsheet action of the heterotic strings considered below takes the
general form [\cHet]:
 $$
\eqalign{{\cal S}=&-{1\over{4\pi \alpha ^{^{\prime }}}}\int d\tau d\sigma
[G_{ab}\partial _\alpha X^a\partial ^\alpha X^b +
B_{ab} \epsilon_{\alpha \beta} \partial _\alpha X^a\partial ^\beta X^b+
\cr
&i\bar \psi^a\gamma ^{-}(\partial _\tau +\partial_\sigma )\psi^a
 +A^I_a \partial _\alpha X^a\partial ^\alpha X^I+g_{IJ}\partial
_\alpha  X^I\partial ^\alpha X^J]\cr} \equn\put\one
$$
where the $\sigma$ and $\tau$ label the worldsheet
coordinates\footnote{$^*$} {in the following we use $ a\in\{
1,\ldots,10\}$, $\mu=1,...,4$, $i=5,\ldots,10$ and $I=1,\ldots
,16$}, $X^a(\tau ,\sigma )$ are the $2d$ scalars
describing the ten-dimensional target space, $\psi ^a (\tau -\sigma )$
are their right handed fermionic superpartners and the left handed
bosons $X^I(\tau +\sigma )$ describe the gauge lattice
$\Gamma_{16}$ of $E_8\times E_8$ or $Spin(32) / {\bf Z}_{2}$. In
{\one} the metric of the lattice $\Gamma_{16}$ is denoted by $g_{IJ}$,
while $G_{ab}$  is the gravitational and $B_{ab}$ is
the the antisymmetric background fields.

The simplest $4d$ string models are obtained through the toroidal
compactification of the six internal coordinates $X^i(\tau ,\sigma
)$ on a torus $ T^6 = {{\bf R}^6 /{2 \pi \Lambda}}$
with $\Lambda$ a $6$-dimensional lattice with basis $\{{\bf
e}_{i}\}$. This compactification is achieved through the
identification:
$$
X^{a}= X^{a}+ 2{\pi}n^i ({\bf e}_{i})^a \;\ n^i\in {\bf Z}
\equn\put\nine
$$

It is useful to define the metric $G_{ij} =  {\bf
e}_{i}\cdot {\bf e}_{j}$ of the internal lattice $\Lambda$ and its
inverse $G^{ij}=G^{-1}_{ij}$. The standard basis vectors of the dual
lattice
 $\Lambda^{*}$ of $\Lambda$ are $\{ {\bf e}^{*i}=
G^{ij}{\bf e}_{j}\}$. The gauge lattice $\Gamma_{16}$ is
spanned by the basis vectors ${\bf e}_I$.

 Any state in the Hilbert space is obviously associated with the
set of quantum numbers $(p^{\mu}, m_i, n^i, p^I)$ where $p^{\mu}$,
$\mu = 1,\ldots,4$, is the continuous space-time momentum, $ {\bf
p} =  m_i{\bf e}^{*i} \in \Lambda^{*} \;\ (m_i\in {\bf Z})$  is the
internal momentum, $ {\bf n} = n^i{\bf e}_{i} \in \Lambda \;\ (
n^i\in {\bf Z}) $ is the winding number appearing in {\nine}, and
$P=p^I{\bf e}_I$, $I=1,\ldots,16$ is its gauge internal number.

The background gauge fields $A^I_a$  and
antisymmetric tensor $B_{ij}$ can be written in the
$\Gamma$ lattice frames as $A^I_a=a^I_i ({\bf e}^{*i})_a $ and
$B_{ab}=b_{ij}({\bf e}^{*i})_a ({\bf e}^{*i})_b$. The
corresponding physical quantities are the invariant integrals,
in one to one correspondence with the lattice vectors of the six
dimensional internal lattice which in fact corresponds to the loops
with non trivial homotopy. These integrals, called Wilson
lines, are given by $\int_i dx^a A^I_{a}= 2 \pi
a^I_i$  and the surface integrals $\int_{ij} dx^a \wedge dx^b  B_{ab}=
4 \pi^2 b_{ij}$.

In the following, we restrict ourselves to the case with vanishing
antisymmetric background field $b_{ij}=0$. When the gauge background
fields also vanishes $A^I_a=0$, the internal momenta can be combined
into a vector  $\left(
{\bf p}_{L} ; {\bf p}_{R} \right) =
     \left( p_{L}^{I} {\bf e}_{I} + p_{L}^{i} {\bf e}_{i} ;
       p_{R}^{i} {\bf e}_{i}  \right)=
\left( P  , p_{L}  ; p_{R}   \right)$ with
$$\eqalign{ p_{L}^{I} &= p^{I}, \cr
 p_{L}^{i} &=  {1\over2}m^{i}  + n^i, \cr
 p_{R}^{i} &=  {1\over2}m^{i}  - n^i, \cr}
\equn\put\twelve $$
lying on a Lorentzian, self-dual, even lattice
$\Gamma=\Gamma_{16+d;d}$ spanned by:

$$\eqalign{{\bf k}^{i} &= \left( 0, {1\over2} {\bf e}^{*i};{1\over2}
{\bf e}^{*i}  \right) \cr \ov{{\bf k}}_{i} &= \left( 0, {\bf
e}_{i};- {\bf e}_{i} \right) \cr {\bf l}_{I} &= \left(
{\bf e}_{I}, 0;0 \right). \cr} \equn\put\thirteen
$$

The precise effect of the Wilson lines is obtained by considering
their modification to the internal momenta [\cNar]:

$$\eqalign{ p_{L}^{I} &= p^{I} + A^{I}_{i}n^i \cr
p_{L}^{i} &=  {1\over2}m^{i} - {1\over2} p^{J}A^{Ji} - {1\over4}
A^{Ki}A^{K}_{j}n^{j} + n^i \cr
 p_{R}^{i} &=  {1\over2}m^{i}- {1\over2}
p^{J}A^{Ji} - {1\over4} A^{Ki}A^{K}_{j}n^{j}  - n^i \cr}
\equn\put\mwil $$
which corresponds to acting on the lattice {\thirteen} by a Lorentz
boost represented by the matrix [\cGin]:
$$
\tremat{\delta^{I}_{J}}{{1\over2}A^I_{b}}
{-{1\over2}A^I_{b'}}
{-{1\over2} A^J_a}{\delta_{ab}-{1\over8}
A^{K}_a A^{K}_{b}} {{1\over8}
A^{K}_a A^{K}_{b'}}
{-{1\over2}A^J_{a'}}{-{1\over8}
A^K_{a'} A^{K}_{b}}{\delta_{a'b'}+{1\over8}
A^K_{a'}A^{K}_{b'}}=
exp{1\over2}\tremat{0}{A^I_{b}}{-A^I_{b'}}{-A^J_a}{0}{0}
{-A^J_{a'}}{0}{0}
\equn\put\twenty $$

After this boosting , as it can be read from {\mwil}
some previously massless string states become massive. When these
states are identified with some vector gauge boson fields, the gauge
symmetry is broken [\cNar-\cHig]. Here, the role of the Higgs
field is played by the scalar corresponding to the vertex
operator $\partial _\alpha X^a\partial ^\alpha X^I$. It is obvious
from the form of this operator that the Wilson lines are associated
with the Cartan subalgebra and thus the Higgs field is in the adjoint
representation of the gauge group. As a result the rank of the group
cannot be reduced.

Within these notations, it is easy to implement the
supersymmetry breaking effect. In fact, the Scherk-Schwarz mechanism
can be described by a Lorentzian boost on the vector $(Q^B, p^b_L;
p^{b'}_R)$ of the extended lattice, obtained by adding some
charge generator $Q^A$ to the Narain lattice [\cSSb, \cAnt]:

$${\left(\matrix{{\delta_{AB}}&{-{1\over2}\xi_{Ab} }
&{{1\over2}\xi_{Ab'} }\cr
{{1\over2}\xi^*_{aB} }&{\delta_{ab} -{1\over8}\xi_{Ca} \xi^*_{Cb}}
&{{1\over8}\xi_{Ca} \xi^*_{Cb'}}\cr
{{1\over2}\xi^*_{a'B} }&{-{1\over8}\xi_{Ca'} \xi^*_{Cb}}&
{\delta_{a'b'}+{1\over8}\xi_{Ca'} \xi^*_{Cb'}} \cr }\right)}=
exp {1\over2}\tremat{0}{-\xi_{Ab} }{\xi_{Ab'} } {\xi^*_{aB}
}{0}{0}{\xi^*_{a'B} }{0}{0}  \equn\put\twentyone $$

The combination of the both Wilson lines and Scherk-Schwarz
charge is then equivalent to a boost on the vector $(Q^A, p^I,
p^a_L; p^{a'}_R)$ given by:

$${\left(\matrix{{\delta_{AB}}&0&{-{1\over2}\xi_{Ab} }
&{{1\over2}\xi_{Ab'} }\cr
0&{\delta^{I}_{J}}&{{1\over2}A^I_{b}}
&{-{1\over2}A^I_{b'}}  \cr
{{1\over2}\xi^*_{aB} }&{-{1\over2}A^J_a}&{\delta_{ab}-{1\over
8}\xi_{Ca} \xi^*_{Cb}
 -{1\over8} A^{K}_aA^{K}_{b}}
&{1\over8}{\xi_{Ca} \xi^*_{Cb'}
 +{1\over8}
A^{K}_aA^{K}_{b'}}\cr
{{1\over2}\xi^*_{a'B} }&{-{1\over2}A^J_{a'}}&{- {1\over8}\xi_{Ca'}
\xi^*_{Cb} -{1\over8} A^{K}_{a'} A^{K}_{b}}&
{\delta_{a'b'}+{1\over8}\xi_{Ca'} \xi^*_{Cb'}+{1\over8} A^K_{a'}
A^{K}_{b'}}\cr}\right)} \equn\put\twentytwo$$
which leads to the following spectrum:

$$\eqalign{ Q^{A} &\rightarrow Q^{A} - \xi^{A}_j n^j \cr
 p_{L}^{I} &\rightarrow p^{I} + A^{I}_{j}n^j \cr
 p_{L}^{i} &\rightarrow  {1\over2}(m^{i}+\xi^{*i}_j Q^j
-{1\over2} \xi^i_j \xi^{*j}_k n^k -  p^{J}A^{Ji} - {1\over2}
A^{Ki}A^{K}_{j}n^{j}) + n^i \cr
 p_{R}^{i} &\rightarrow {1\over2}(m^{i}+\xi^{*i}_j Q^j
-{1\over2} \xi^i_j \xi^{*j}_k n^k -  p^{J}A^{Ji} - {1\over2}
A^{Ki}A^{K}_{j}n^{j})  - n^i,\cr}
\equn\put\twentysix $$

It is important to notice that the breaking of gauge symmetry and
supersymmetry commute as they are two similar (but different as
explained below) Lorentz boosts. Then there is no new condition
imposed on the charge $Q^A$. That allows us to study the properties of
the models in their supersymmetric phase. This was expected but it is
not automatic, because going to the non-supersymmetric phase, one
gives a vacuum expectation value for an unphysical (non-BRST invariant)
auxiliary field [\cSSb]. The two phases are not continuously connected
and they do not have the same classical moduli space.

The toroidal compactification models described above lead to a non
realistic massless spectrum as it has an extended $N=4$ supersymmetry
which has no chiral representations. The natural simple candidate to
study the Scherk-Schwarz mechanism are then the orbifold
compactifications which have an $N=1$ space-time supersymmetry
[\corb]. Below, we will restrict our attention to symmetric abelian
$Z_N$ and $Z_N \times Z_M$ orbifolds.

We consider now the toroidal models above where $T^6$ is seen as the
product of three tori  $T^6={{T^2} \times {T^2} \times {T^2}}$
described by  three complexes coordinates $X^{\alpha}$ with $i=4,5,6$
and the corresponding fermions:
$$
f_\alpha =\psi_{3+2\alpha} +i \psi_{4+2\alpha}
\equn\put\fdef
$$
The orbifolds compactifications are obtained by dividing out some
discrete subgroup $G$ of the automorphisms of the Hilbert space. $G$
includes the space lattice twists corresponding to the elements
$g=(\theta,v)$ which form the space group. They define the action of
$G$ on the quantum numbers $p_{L,i}, p_{R,i}$ of any state. The pure
translations $(1,v)$ form the lattice $\Lambda$  and the rotations
$\theta$ form the point group  $P$. To preserve $N=1$ supersymmetry
$P$ must be a discrete crystallographic  subgroup of $SU(3)$. We are
interested in the case where $P={\bf Z}_N$ (or $P={\bf Z}_N \times
{\bf Z}_M$) is  generated by a rotation ${\theta}$ of order $N$ (or
rotations ${\theta}, \omega$ of orders $N$ and $M$ respectively) . The
elements of the space group have the form $g=({\theta}^j,v)$ with
$j=0,1,...,N-1$ (or $g=({\theta}^j\omega^k,v)$ with $j=0,1,...,N-1$,
and $k=0,1,...,M-1$).

The physical Hilbert space for a string propagating on an
orbifold consists in different sectors. There are the twisted sectors
containing states not present in the toroidal compactification. These
states satisfy the boundary condition:
$$
X({\sigma + 2\pi}) = \theta^k X({\sigma}) + {2 \pi v} \qquad
k=1,...,N-1
\equn\put\twentynine
$$
They don't have internal momenta, so they don't feel the supersymmetry
breaking mechanism at the tree level. There is also an untwisted sector
which is obtained from the Hilbert space of a string  propagating on a
torus by projecting on invariant states under the action of $G$. These
states are defined by the closed string boundary conditions:
$$
X({\sigma + 2\pi}) = X({\sigma}) + {2 \pi v}
\equn\put\twentyeight
$$
where $v$ are the lattice $\Lambda$ vectors. Their mass spectrum is
determined by the associated internal momenta through:
$$
{1\over 4}m_L^2={1\over 4}m_R^2=N_R + {1\over 2} {\bf p}_R^2 ,
\equn\put\thirty
$$
with ${\bf p}_R$ defined in {\twentysix} and $N_R$ is the oscillator
number. The requirement that orbifold projection, gauge symmetry and
supersymmetry breaking commute, restricts the allowed Wilson lines and
the charges $Q^A$.

 Obviously, the two effects of gauge symmetry and supersymmetry
breaking do not mix. We can then study the two effects separately.
Let's first discuss the possible charges that could be used for
supersymmetry breaking.

The charge $Q^A$ can be written as:
$$ Q^A= \oint J^A
\equn\put\charge$$
where $J^A$ is a $U(1)$ current which should satisfy the following
requirements:

i) $J^A$ must obviously be a dimension one conformal operator
satisfying the $U(1)$ algebra.

ii) $J^A$ shouldn't commute with the $2d$ supercurrent so
that it gives a charge for the gravitino but not to the graviton and
gauge bosons which are usually in the untwisted sector.

The current $J^A$ has then the form of a bilinear in the fermions:
$$
J^A = C_{\alpha \beta} \psi^\alpha \psi^\beta
\equn\put\charge
$$

In orbifold compactifications, the condition that $Q^A$ is associated
with some particular direction $A$ means that it should have the same
transformation under the orbifold group than the corresponding
coordinate $\partial X^A$. This requirement is very strong as it leaves
only few possible $U(1)$ currents.

We have listed the different currents we found in the tables 1 and 2
for the cases of $Z_N$ and $Z_N\times Z_M$ orbifolds respectively. For
the orbifolds ${\bf Z}_4$ and ${\bf Z}_2 \times {\bf Z}_2$ the $U(1)$
charges were already given in [\cAnt]. We have focused on orbifolds
where only one dimension (${\bf Z}_2$ case) or two dimensions are
large. For example, this excludes ${\bf Z}_7$ orbifold which needs the
six internal dimensions to be of the same size. For the orbifolds ${\bf
Z}_3$, ${\bf Z}_6$, ${\bf Z}_8$ and ${\bf Z}_{12}$ as well as ${\bf
Z}_2 \times {\bf Z}_6 $, ${\bf Z}_3 \times {\bf Z}_3$,  ${\bf Z}_3
\times {\bf Z}_6$, ${\bf Z}_4 \times {\bf Z}_4 $ and ${\bf Z}_6
\times {\bf Z}_6$ we have not found charges allowing to implement the
Scherk-Schwarz mechanism.

Only two orbifolds ${\bf Z}_4$ and ${\bf Z}'_6 \equiv {\bf Z}_2
\times {\bf Z}_3$ have charges associated with $N=4$ sectors where
the threshold corrections vanish. The other orbifolds lead to light
KK-states in $N=2$ multiplets. In this case, the one loop threshold
depending on the value of the large radius is not automatically
vanishing. One would have then to chose the particle content as
KK-excitations to get vanishing $\beta$-functions [\cAnt].

We will now discuss some relevant properties of
the Kaluza-Klein spectrum. For this we need to make the size of the
internal radii explicit. This is done through the following
redefinitions of the lattice vectors:
$$
{\bf e}_{i}  \rightarrow R_i {\bf e}_i\;\;\;\;\;\;\;\;\;\;\;\;
        {\bf e}^{*i}  \rightarrow {1\over R_i} {\bf e}^{*i},
\;\;\;\;\;\;\;\;\;\;\;\; G_{ij} \rightarrow \delta_{ij}\ \,
\equn\put\thirtyone$$
The vector ${\bf p}_R$  appearing in the mass formula
{\thirty} takes then the well known form:

$$ {\bf p}_R = ( m_i - a^I_i (p^I -{1\over2} a^I_j n^j)) {{\bf e}^{*i}
\over {2R_i} }- n^i R_i {\bf e}_i,
\equn\put\thirtytwo
$$

 We also make the rescaling $\xi^A_i \rightarrow
\xi^A_i/R_i$. A well known result is that in
superstring theory these new parameters $\xi^A_i$ are not continuous,
but can only take discrete values of order one [\cPre, \cSSb]. The
corresponding formula in the presence of the supersymmetry breaking is
then:

$$ {\bf p}_R = ( m_i - a^I_i (p^I -{1\over2} a^I_j n^j)+\xi^*_{ij} Q^j
- {{\xi_{ki} \xi^*_{kj}}\over 2} n^j) {{\bf e}^{*i} \over {2R_i} } +
n^i R_i  {\bf e}_i,
\equn\put\thirtythree$$

while:
$$ {\bf p}_L = ( m_i - a^I_i (p^I -{1\over2} a^I_j n^j)+\xi^*_{ij}
Q^j - {{\xi_{ki} \xi^*_{kj}}\over 2} n^j) {{\bf e}^{*i} \over {2R_i} }
- n^i R_i  {\bf e}_i.
\equn\put\thirtyfour$$

In the above formula the charge $Q^j$ takes integer and
fractional values for the bosons and fermions respectively. The
requirement that the orbifold projection and gauge symmetry breaking
commute imposes a condition on the allowed Wilson lines
[\corb-\cWil]:  $$  N{\bf a} \in \Gamma \equn\put\twentyseven$$
Moreover, the invariance under $e_i\rightarrow e_{i+1}$, if part of
the point group, implies that $a_i=a_{i+1}$ and thus reduces the
maximum number of independent discrete Wilson lines.

Although the formula above in the absence of supersymmetry breaking
are well known, their implications for a large internal dimension have
not been studied.

Notice first that in the gauge symmetry breaking process, the states
acquire masses  inversely proportional to the radius of the torus
corresponding to the Wilson line. The massless states (in the
supersymmetric phase) can easily be seen to correspond to Wilson lines
singlets: $aP \in {\bf Z}$. While we see that a Wilson line
associated to the torus with the large radius used to break
supersymmetry will lead to a mass of the order of hundreds GeV or TeV
to the projected states. In particular, if some states have $0 < |aP|
< 1$ as it is often the case, then the corresponding states will
have masses smaller than the KK excitations of the states, with
different gauge quantum numbers, present at the massless level. This
also implies that the minimal light KK states are obtained when all the
Wilson lines have to be associated only to the other small tori. Such
a minimally requirement would also automatically avoid the presence of
some massive new vector bosons that could mediate new dangerous
interactions.

The appearance in the untwisted sector of massless vector bosons
in the adjoint representations  implies the presence of their KK
excitations. As shown in [\cAB], the left moving Hilbert
states, carrying gauge quantum numbers, transform under the orbifold
group by phases that can be compensated by lattice phases. The
fermions and scalars partners of the vector representations can then
be obtained by exchanging the $\partial X^{\mu}$s by any right handed
oscillator are present in the massive KK spectrum with the same mass.
If all the three internal two-dimensional tori are untwisted
simultaneously, the above representations generate $N=4$ multiplets in
the adjoint representation. If not all of the internal coordinates
are untwisted we obtain $N=2$ multiplets.

The formula {\thirtythree} shows that all the states carrying
the same gauge internal momenta have the same masses. In particular,
this implies that all the $N=2$ and $N=4$ multiplets get
projected by the gauge symmetry breaking and only $N=4$ (or
$N=2$) excitations of massless untwisted states are present among the
light KK states.

To illustrate the above point let's consider the simplest case of the
presence of a massless untwisted state charged only under one
$U(1)$. This state  of charge
$Q$ and denoted by $S_0$ (index $0$ for massless)  has a tower of KK
excitations $S_n$ of masses $n/R$. They are supposed to come in
massive $N=4$ multiplets containing 5 scalars, 4 fermions and one
vector. The vector excitation has also the same charge $Q$ under
$U(1)$, so it must be part of the adjoint representation of a
non-abelian group.  Moreover, as the $U(1)$ gauge group is unbroken,
it must be anomaly free.  In the simplest case, it is sufficient to
have in addition a massless state ${\bar S_0}$ of charge ${-Q}$ and
the symmetry group at the massive level will be $SU(2)$. Notice that
as the (unbroken) gauge symmetry and the massless (twisted)
representations are not given by the extended group, the couplings
between the massless and KK states are more complicated [\cAB]. The
generalization of these arguments to the presence of other
representations is straightforward.  For example, by applying them one
could obtain the light KK spectrum when the untwisted sector contain
chiral supermultiplets transforming under $SU(N)\times SU(M)\times
U(1)$ as $(1,1)_Q$, $(N,1)_Q$, $(1,M)_Q$ or $(N,M)_Q$.

Finally, we have to deal with the effect of reducing the
rank of the gauge group on the Kaluza-Klein excitations. The Higgs
mechanism through discreet Wilson lines described above doesn't reduce
the rank of the gauge group. To reduce the rank, there are usually two
mechanism used. The first is to embed the Wilson lines in
the gauge group as automorphism of the $\Gamma_{16}$ lattice [\crank].
This corresponds to the case where the orbifold action
on the gauge lattice is described by a rotation ${\Theta}\neq 0$.
In this case some Cartan generators of the gauge group are not
associated with a root of $\Gamma_{16}$, but with an invariant
combination of winding states. In the case where some components of
the Wilson line are rotated by ${\Theta}$, the
projection on Wilson line singlets projects out, in general, the
Cartan generators which have the form of invariant combination of
winding states. As this projection is at the level of the gauge
lattice state, which is the same for all the KK excitations of
the gauge boson, all the KK tower is projected out. Both the rank of
the gauge symmetry group and the rank of the symmetry group of the
KK excitations are reduced simultaneously. Notice that if the Wilson
line is associated with the dimension with large size then the
projected states are very light.

Another often used way to decrease the rank is to exploit the fact
that in these compactifications one $U(1)$ gauge factor is usually
anomalous and develops a  Fayet-Iliopoulos D-term [\cFI],
meaning that the classical vacuum is ill defined. This $D$-term is
canceled by giving a ``small" vev (of the order of the coupling
constant) to a massless state (modulus) charged under this $U(1)$
[\cdtrm]. The $U(1)$ factor as well as the gauge factors under which
this state is charged are all broken. From the effective field theory
point of view, the KK excitations couple also to the shifted scalars
and all the tower of KK states get large additional masses. At the
level of the string formulation, this effect is similar to the
Fishler-Susskind mechanism and corresponds to cancellation of tadpoles
between one and two loops effects and is beyond the $2d$ conformal
field theory techniques of string perturbation theory.

It would be useful to exhibit some  explicit semi-realistic models
where the perturbative mechanism of supersymmetry breaking could be
studied. Other important questions like a solution for the
cosmological constant problem or the computation of higher-loops
threshold corrections in the broken phase remain still open.

\vskip 0.5cm
\noindent{\bf Acknowledgments}
\vskip.5cm
I wish to thank I. Antoniadis, E. Gava, K.S. Narain and A. Sen for
useful discussions.

\vskip 1.5cm
\centerline{\bf REFERENCES}
\vskip 0.5cm

\parskip=-3 pt

\item{[{\cPre}]} R. Rohm and E. Witten, Ann. Phys. 170 (1986) 454; T.
Banks and L. Dixon, {\np} {\bf B307} (1988) 93; I. Antoniadis, C.
Bachas, D. Lewellen and T. Tomaras, {\pl} {\bf 207B} (1988) 441; S.P.
de Alwis, J. Polchinski and R. Schimmrigk, {\pl} {\bf 218B} (1989)
449. \hfill\break

\item{[{\cSSb}]}
R. Rohm, {\np} {\bf B237} (1984) 553; C.~Kounnas and M.~Porrati, {\np}
{\bf B310 } (1988) 355; S.~Ferrara, C.~Kounnas, M.~Porrati and
F.~Zwirner, {\np} {\bf B318} (1989) 75; C. Kounnas and B. Rostand,
{\np} {\bf B341} (1990) 641.
\hfill\break

\item{[{\cBac}]}  C. Bachas, preprint  CPTH-R349-0395
hep-th/9503030 (1995).
\hfill\break

\item{[{\cSS}]}
J.~Scherk and J.H.~Schwarz, {\pl} {\bf B82} (1979) 60 and {\np}
{\bf B153} (1979) 61; E.~Cremmer, J.~Scherk and J.H.~Schwarz,{\pl} {\bf
B84} (1979) 83; P.~Fayet, {\pl} {\bf B159} (1985) 121 and {\np} {\bf
B263} (1986) 649;Proc. 2nd Nobel Symposium on El. Part. Physics at
Marstrand (Sweden, 1986), Physica Scripta T15 (1987) 46
\hfill\break

\item{[{\cAnt}]} I. Antoniadis, {\pl} {\bf 246B} (1990) 377; Proc.
PASCOS-91 Symposium, Boston 1991 (World Scientific, Singapore)
p.718.
\hfill\break

\item{[{\cAqm}]} I. Antoniadis, C. Mu\~noz and M. Quir\'os, {\np} {\bf
B397} (1993) 515.\hfill\break

\item{[{\cAB}]} I. Antoniadis and K. Benakli {\pl} {\bf 326B} (1994)
69.
\hfill\break

\item{[{\cABQ}]} I. Antoniadis, K. Benakli and M. Quir\'os  {\pl} {\bf
331B} (1994) 313.\hfill\break

\item{[{\cABQ}]} K. Benakli, in preparation.\hfill\break

\item{[{\cHet}]} D.J. Gross, J.A. Harvey, E. Martinec and R.
Rohm, {\np} {\bf B256} (1985) 253; {\bf B267} (1986) 75.
\hfill\break

\item{[{\cNar}]} K.S. Narain, {\pl} {\bf B169} (1986) 41; K.S. Narain,
M.H. Sarmadi and E. Witten {\np} {\bf B279} (1987) 369.
\hfill\break

\item{[{\cGin}]} P. Ginsparg, {\pr} {\bf D35} (1987) 648.
\hfill\break

\item{[{\cHig}]} S. Ferrara, C. Kounnas and M. Porrati, {\np} {\bf
B304} (1988) 500; I. Antoniadis, C. Bachas  and C. Kounnas, {\pl} {\bf
200B} (1988) 297.
\hfill\break

\item{[{\corb}]} L.J. Dixon, J. Harvey, C. Vafa and E. Witten, {\np}
{\bf B261} (1985) 678; {\bf B274} (1986) 285.
\hfill\break

\item{[{\cWil}]}  L.E. Ib\'a\~nez, H. P. Nilles, F. Quevedo,
{\pl}  {\bf 187B} (1987) 25.
\hfill\break

\item{[{\crank}]} L.E. Ib\'a\~nez, H. P. Nilles, F. Quevedo,
{\pl}  {\bf 192B} (1987) 332.
\hfill\break

\item{[{\cFI}]} P. Fayet and J. Iliopoulos, {\pl}  {\bf 51B} (1974)
461.
\hfill\break

\item{[{\cdtrm}]} M. Dine, N. Seiberg and E. Witten {\np}
{\bf B289} (1987) 585; J.J. Atick, L. Dixon and A. Sen, {\np}
{\bf B292} (1987) 109; M. Dine, I. Ichinose and N. Seiberg, {\np}
{\bf B293} (1987) 253.
\hfill\break

\vfill\eject
\vbox {\tabskip=0pt \offinterlineskip\def\tablerule{\noalign{\hrule}}
\def\tv{\vrule height 20pt depth 5pt}\halign to 13cm {\tabskip=0pt
plus 20mm \tv\hfill\quad#\qquad\hfill &\tv\hfill\quad#\qquad\hfill
&\tv\hfill\quad# \qquad\hfill &\tv\hfill\quad#\quad\hfill
&\tv#\tabskip=0pt\cr\tablerule  ${\bf {\bf Z}}_N$  &  shift &
$(N;A)$ & $J^A$
&\cr\tablerule
${\bf {\bf Z}}_4$&${1\over 4}(1,1,-2)$ &
$(4;1,2)$&$J^1={1\over \sqrt{2}}(f^*_1 Re(f_3) + f_1 \psi )$  &
\tabskip=0pt\cr & &$(2;3)$&$ {1\over \sqrt{2}}(f^*_1 f^*_2
+ f_1 f_2 )$ &
\tabskip=0pt\cr & & &$\psi Re(f_3)$ &
\tabskip=0pt\cr
${\bf {\bf Z}}'_6$&${1\over 6}(1,2,-3)$&
$(4;1)$&${1\over \sqrt{3}}(f^*_1 f_2 + f_1 \psi+ f^*_2
Re(f_3))$  &
\tabskip=0pt\cr & &$(2;3)$ &${1\over \sqrt{2}}(f^*_1 f^*_2
+ f_1 f_2 )$ &
\tabskip=0pt\cr & & &$\psi Re(f_3)$ &
\tabskip=0pt\cr
 ${\bf {\bf Z}}'_8$&${1\over 8}(1,3,-4)$&
$(2;3)$&${1\over \sqrt{2}}(f^*_1 f^*_2 + f_1 f_2 )$ &
\tabskip=0pt\cr & & &$\psi Re(f_3)$ &
\tabskip=0pt\cr
${\bf {\bf Z}}'_{12}$&${1\over 12}(1,5,-6)$&
$(2;3)$&${1\over \sqrt{2}}(f^*_1 f^*_2
+ f_1 f_2 )$ &
\tabskip=0pt\cr & &&$\psi Re(f_3)$ &
\tabskip=0pt\cr\tablerule}}

\vskip 1.0truecm
\centerline{\bf Table 1.}
\vskip 0.5truecm
Currents $J^A$ that can be used for breaking supersymmetry in ${\bf
{\bf Z}}_N$ orbifolds. We specify the sector $N=2$ or 4 and the
corresponding directions $A$. The currents are given up to trivial
redefinitions, as the change to another equivalent direction $A$.
$\psi$ is one of the $\psi^\mu$s.

\vfill\eject
\vbox {\tabskip=0pt \offinterlineskip\def\tablerule{\noalign{\hrule}}
\def\tv{\vrule height 20pt depth 5pt}\halign to 14cm {\tabskip=0pt
plus 20mm \tv\hfill\quad#\quad\hfill &\tv\hfill\quad#\quad\hfill
&\tv\hfill\quad# \quad\hfill &\tv\hfill\quad#\quad\hfill
&\tv#\tabskip=0pt\cr\tablerule  ${\bf Z}_N\times {\bf Z}_M$  & shift
&$(N;A)$& $J^A$  &\cr\tablerule
${\bf Z}_2 \times {\bf Z}_2 $&$({1\over 2},0,-{1\over
2})\times (0,{1\over 2},{1\over 2})$  &$(2;1,2,3)$ &$J^1=\psi
Re(f_1)$ &
\tabskip=0pt\cr & & &$Re(f_2) Re(f_3)$ &
\tabskip=0pt\cr
$ {\bf Z}_2\times {\bf Z}_3 $&$({1\over 2},0,-{1\over 2})\times (0,{1\over
3},-{1\over 3})$& $(4;3)$&${1\over \sqrt{3}}(f^*_3 f_2 +
f_3 \psi+ f^*_2 Re(f_1))$  &
\tabskip=0pt\cr & &$(2;1)$ &${1\over \sqrt{2}}(f^*_3 f^*_2
+ f_3 f_2 )$ &
\tabskip=0pt\cr & & &$\psi Re(f_3)$ &
\tabskip=0pt\cr
 $ {\bf Z}_2 \times {\bf Z}_4 $&$({1\over 2},0,-{1\over 2})\times
(0,{1\over 4},-{1\over 4})$&
$(2;1)$&${1\over \sqrt{2}}(f^*_3 f^*_2 + f_3 f_2 )$ &
\tabskip=0pt\cr & & &$\psi Re(f_1)$ &
\tabskip=0pt\cr & &$(2;2,3)$ &$J^2={1\over \sqrt{2}}(f_2
\psi + f^*_2 Re(f_1) )$ &
\tabskip=0pt\cr
$ {\bf Z}_2 \times {\bf Z}'_6$& $({1\over 2},0,-{1\over 2})\times
(0,{1\over
6},-{1\over 6})$&
$(2;1)$&${1\over \sqrt{2}}(f^*_3 f^*_2
+ f_3 f_2 )$ &
\tabskip=0pt\cr & &&$\psi Re(f_1)$ &
\tabskip=0pt\cr\tablerule}}

\vskip 1.0truecm
\centerline{\bf Table 2.}
\vskip 0.5truecm
Currents $J^A$ that can be used for breaking supersymmetry in ${\bf
{\bf {\bf Z}}}_N$ orbifolds. We specify the sector $N=2$ or 4 and the
corresponding directions $A$. The currents are given up to trivial
redefinitions, as the change to another equivalent direction $A$.
$\psi$ is one of the $\psi^\mu$s.
  \end